# A Cloud Platform-as-a-Service for Multimedia Conferencing Service Provisioning

Ahmad F. B. Alam[†], Abbas Soltanian[†], Sami Yangui[†], Mohammad A. Salahuddin[†‡],
Roch Glitho[†] and Halima Elbiaze[‡]

[†]Concordia University, Montreal, Quebec, Canada, [‡]Université du Québec À Montréal, Montreal, Quebec, Canada
{a_binala, ab_solta, s_yangui, glitho}@encs.concordia.ca, mohammad.salahuddin@ieee.org, elbiaze.halima@uqam.ca

*Abstract*—Multimedia conferencing is the real-time exchange of multimedia content between multiple parties. It is the basis of a wide range of applications (e.g., multimedia multiplayer game). Cloud-based provisioning of the conferencing services on which these applications rely will bring benefits, such as easy service provisioning and elastic scalability. However, it remains a big challenge. This paper proposes a PaaS for conferencing service provisioning. The proposed PaaS is based on a business model from the state of the art. It relies on conferencing IaaSs that, instead of VMs, offer conferencing substrates (e.g., dial-in signaling, video mixer and audio mixer). The PaaS enables composition of new conferences from substrates on the fly. This has been prototyped in this paper and, in order to evaluate it, a conferencing IaaS is also implemented. Performance measurements are also made.

## I. Introduction

Cloud computing is a paradigm for swiftly provisioning a shared pool of configurable resources (e.g., storage, network, application and services) on demand. It has three key facets: Software-as-a-Service (SaaS), Platform-as-a-Service (PaaS), and Infrastructure-as-a-Service (IaaS) [1]. It provides several benefits, such as rapid provisioning of services, scalability and elasticity. Multimedia conferencing is the conversational exchange of media content (e.g., voice, video and text) between multiple parties [2]. It is an important component of *conferencing applications* (e.g., audio/video conference, massively multiplayer online games).

For cost efficiency purpose, developers of conferencing applications can use *conferencing services* (e.g., dial-in video conference and dial-out audio conference with floor control) offered by third parties. Such services could be provisioned as SaaS by third party conferencing service providers using PaaS. *Conferencing service provisioning* refers to the entire life-cycle of the conferencing service, i.e. development, deployment and management [3]. Provisioning conferencing services in the cloud is quite challenging. A challenge, for instance, is the necessity for conferencing service developers to master low-level details of conferencing technologies, protocols and their interactions. Yet, another challenge is that the provisioned conferences need to scale elastically to accommodate a fluctuating number of participants. Unfortunately, existing PaaS solutions do not address these challenges. This paper proposes an innovative PaaS to tackle them.

The proposed PaaS is based on the business model in [2], which proposes six roles: Connectivity provider, broker, conferencing substrate providers, conferencing infrastructure providers, conferencing platform providers and conferencing service providers. This paper focuses on conferencing service providers, conferencing platform providers and conferencing infrastructure providers. Moreover, it is assumed that the substrate provider plays the role of the conferencing infrastructure provider as well. The proposed PaaS provides conferencing service providers, who are experienced in programming, with high-level interfaces to hide the internal complexities of conferencing. Besides, it composes on the fly conferencing building blocks entitled substrates (e.g., dial-in signaling, video mixer and audio mixer) into full-fledged conferences.

The rest of the paper is organized as follows: Section II introduces a motivating scenario, derives the requirements on conferencing PaaS and reviews related work. Section III describes the proposed overall architecture. Section IV presents the software architecture, prototype and experimental results. Section V concludes the paper.

## II. Motivation, requirements and related work

### A. Motivating Scenario

Fig. 1 depicts the motivating scenario. It includes three example conferencing applications – (i) a game using dial-in audio conferencing, (ii) a distance learning program using dial-out audio conferencing, and (iii) a plain conferencing application offering dial-out video conference with floor control. These conferencing applications use conferencing services offered as SaaS by conferencing service providers. One service provider offers conferencing service A that supports both dial-in and dial-out audio conferences. The distance learning and the game applications consume service A. Another provider offers dial-out video conference with floor control (service B), used by the plain conferencing application.

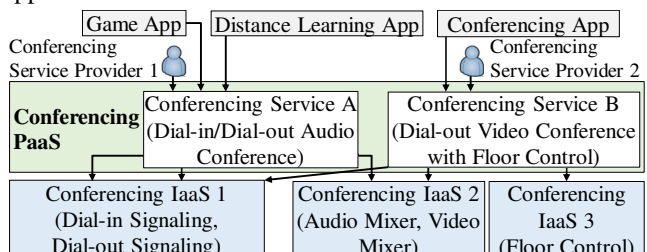

Fig. 1. Conferencing Service Provisioning in the Cloud

The conferencing services create new conferences when they receive corresponding requests from conferencing applications. For example, service A creates dial-in audio conference and dial-out audio conference when it receives requests from the distance learning and the game applications, respectively. The PaaS composes these conferences from the required substrates offered by conferencing IaaSs. It is assumed that the PaaS has prior knowledge of the existing conferencing IaaSs and their offered substrates. During a conference, as the players join and leave, the PaaS scales the conference in an elastic manner.

### B. Requirements

The following four requirements are derived:

*1) High-level Interfaces for Service Providers:* The conferencing PaaS interfaces should enable the service providers to provision new services without having to deal with the complexities of conferencing components and their interactions. The interfaces should also be flexible enough for creating complex and novel conferencing services (e.g., a dial-in video conference with five minutes of chat per hour).

*2) Composition of Conferences from Substrates:* When a conferencing service receives a request to create a new conference, the PaaS should determine necessary substrates, select appropriate conferencing IaaSs providing those substrates and then compose the requested conference from the selected substrates.

*3) Elastic Scalability:* The conferencing PaaS, in collaboration with the conferencing IaaSs, should scale the ongoing conferences in response to the fluctuating number of participants. This allows the PaaS to gain cost efficiency and to follow the pay-per-use principle.

*4) Quality of Service:* Meeting Quality of Service (QoS) requirements, such as latency, jitter and throughput, is critical as conferencing services are real-time. This paper focuses on the latency of operations performed during a conference (e.g., participant joining and setting floor chair).

### C. Related Work

The cloud-based conferencing architectures and the existing PaaS are reviewed below.

*1) Cloud-based Conferencing Architectures*

A cloud-based framework for conferencing service provisioning is proposed in [4]. It offers conferencing services as SaaS, while using a conventional PaaS for deployment and execution. It does not provide high-level interfaces to service providers. It does not address conference scalability and QoS requirements, either. An approach for providing video conference as a Web service is presented in [5]. To help the conferencing application developers, this paper proposes a set of high-level SaaS to be offered by conferencing service providers. However, it does not address how service providers could provision these SaaS. Neither does it discuss conference composition, elastic scalability and QoS.

A cloud infrastructure, proposed in [6], relies on conferencing substrates and can scale elastically. It also proposes PaaS/IaaS interfaces rooted in substrates. These characteristics make the infrastructure suitable for use by a conferencing PaaS. However, the PaaS-level issues including the interfaces for service providers and the composition of conference substrates are not taken into account. Neither do they provide QoS measurements for conference runtime operations. There has been other research works in the literature, such as [7], [8], [9], which address specific problems of cloud-based conferencing – for instance, inter-datacenter network utilization, media mixing and transcoding. While they focus on how conferencing components can efficiently utilize the cloud, they do not address conferencing service provisioning.

*2) Existing PaaS Solutions*

Aneka [10] and Cloud Foundry [11], the two representatives of PaaS are evaluated. Aneka provides high-level interfaces and supports elastic scalability, specifically for distributed application provisioning. Nonetheless, the interfaces are not suitable for conferencing service providers. Cloud Foundry provides no interfaces for conferencing service provisioning. It supports scaling of application instances but does not address elastic scaling of conferences. Neither addresses conference composition and QoS.

### III. PROPOSED ARCHITECTURE FOR CONFERENCING PaaS

The architectural principles are first presented. Then, the architectural components and service development APIs are discussed in detail, followed by an illustrative scenario.

### A. Architectural Principles

Two widely used compositional approaches are orchestration and choreography [12]. The former allows a central entity to control the component services and their interactions. The latter allows the component services to collaborate in a decentralized manner. The first principle of our architecture is to adopt the orchestration approach for the substrate composition because it provides PaaS with a greater control on the substrates and their interactions. The second principle is to use high-level PaaS/IaaS interfaces rooted in substrates. It contributes to easy conference composition from substrates. This principle also enables PaaS to request IaaSs for scaling conferences in terms of conference size, instead of VM resources. The third principle is to extend the existing PaaS architectures. This allows us to reuse the existing PaaS for the conferencing PaaS implementation.

### B. Overall Architecture

The proposed architecture consists of a repository and five components, as shown in Fig. 2. These components deal with three key facets: (i) Conferencing services, (ii) conferences and (iii) substrate information.

*1) Components Related to Conferencing Services*

This facet includes service development, deployment and management. *Conferencing PaaS GUIs and APIs* component provides tools for the conferencing service providers. For easy development, service providers use high-level Service Development APIs (c.f. Section III.C), which is novel in this architecture. They also use GUI for service deployment and management, such as starting, updating and stopping services. This component satisfies the requirement of high-level interfaces.

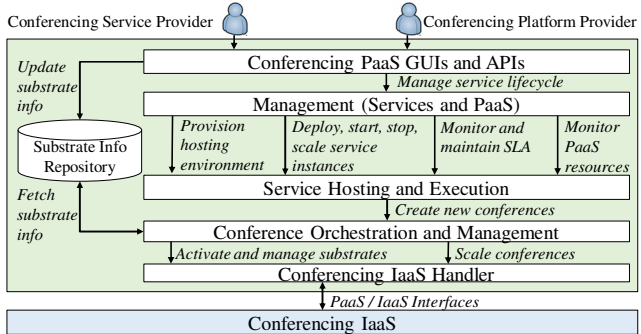

Fig. 2. Overall Architecture of Conferencing PaaS

*Management (Services and PaaS)* component manages the conferencing services and monitors their QoS and SLAs. *Service Hosting and Execution* component hosts the conferencing services. It allocates necessary PaaS resources (e.g., server runtime and database drivers) and prepares execution environment before hosting.

*Management (Services and PaaS)* receives request from the conferencing PaaS GUI for service deployment and management. It deploys and executes services in *Service Hosting and Execution* component and manages them during execution. Note that *Conferencing PaaS GUIs and APIs* is an extension of application provisioning front-end, available in regular PaaS architectures. *Management (Services and PaaS)* and *Service Hosting and Execution* components are reused from conventional PaaS architectures.

2) *Components Related to Conferences*

This facet concerns conference composition and management of created conferences including elastic scaling. *Conference Orchestration and Management* component creates and manages conferences. More explicitly, it performs the following five tasks. First, it determines the necessary substrate types and their associated requirements by using, for instance, syntactic matching with the categorized API parameters (c.f. Section III.C). Second, given the requirements of a substrate, it selects the most suitable conferencing IaaS, by using an algorithm. Existing algorithms for cloud service selection, such as [13], can be reused in this context. Third, it orchestrates conferences from substrates and executes them. Note that conferences are executed in this component. In contrast, the conferencing services that create conferences are executed in the Service Hosting and Execution component. We assume that conferencing IaaSs expose substrates as RESTful web services as in [6]. Therefore, existing approaches and techniques for RESTful web service orchestration, such as [14], can be reused. Fourth, it manages the composed conferences. For example, it can add or remove video from a conference. Fifth, it monitors the current size of each running conference to make decisions about scaling. If needed, it requests conferencing IaaSs to scale in terms of conference size. However, this decision-making process requires new conference scaling algorithms. This component, along with conferencing IaaS, satisfies the requirement of elastic scalability. It also meets the requirement of composing conferences from substrates. *Conferencing IaaS Handler* component handles all communications between the conferencing PaaS and the conferencing IaaSs. It realizes the high-level conferencing PaaS/IaaS interfaces proposed in [6], which is reused in this work.

*Conference Orchestration and Management* receives requests from conferencing services. Based on the requests received (e.g., create a conference and stop a conference), it takes actions and communicates with IaaSs via *Conferencing IaaS Handler*. Note that *Conference Orchestration and Management* is a novel component while *Conferencing IaaS Handler* is an extension of IaaS communication component in conventional PaaS architectures.

3) *Components Related to Substrate Information*

To select the best conferencing IaaS for a given substrate, PaaS needs certain information about that substrate, such as substrate type, price, SLA and QoS. Conferencing PaaS provider uses a GUI in *Conferencing PaaS GUIs and APIs* component to manage (e.g., add, remove, update) such information of the substrates. The information is stored in the *Substrate Information Repository*.

C. *Conferencing Service Development APIs*

Three principles are followed to design the proposed APIs. The first principle is leveraging basic conferencing concepts (e.g., conference, participant, media and floor) in the API design. This helps in achieving an abstraction level higher than conferencing components (e.g., signaling, media mixer and media transcoder) and their complex interactions. The second principle is categorizing API parameters, which helps service providers to easily understand a conference's mandatory and optional aspects, required API parameters for each aspect and dependencies among parameters. The third principle is the use of RESTful design. It is standard-based, lightweight and flexible for data representation, allowing us to describe the APIs in a generic way.

Table I delineates four API examples. It shows some of the REST resources along with an example operation for each. The request parameters and the response contents are also listed. The categorization of API parameters is shown in table II. This table highlights that a service provider has to specify one conference model, at least one media and the conferencing technology. It also shows the conditional dependencies of parameters. For example, for WebRTC-based conferencing, signaling protocol must be specified. The parameters service providers can change during runtime are italicized.

D. *Illustrative Scenario*

The illustrative scenario consists of a game application where players can talk for unlimited time but can have private text chat for only 5 minutes per hour, a service provider offering dial-in audio conferencing service with text chat for limited time and a conferencing PaaS that subscribes to three conferencing IaaSs. IaaS A and B offer dial-in signaling and audio mixer substrates; IaaS C offers an instant messaging substrate. The scenario shows how the conferencing PaaS creates a conference when the game application sends requests to the service. It also illustrates how APIs are used by service providers.

Fig. 3 depicts the interactions. For brevity, the game application is omitted in the figure. When the service receives a request from the game application for creating a conference, it invokes the create conference API. API handling is delegated to *Conference Orchestration and Management*, which determines necessary substrates and selects appropriate IaaSs. It is assumed that it selects IaaS A for dial-in signaling and IaaS B for audio mixer substrates. Next, it requests IaaSs, via *Conferencing IaaS Handler*, to activate the substrates. Interactions for substrate activation are not shown. After activation, *Conference Orchestration and Management* orchestrates a new dial-in audio conference from substrates and then executes it. The orchestrated conference represents a full-fledged conference. It creates individual conferences on the substrates it is composed of. Finally, the ID of the full-fledged conference is returned to the game.

TABLE I. EXAMPLES OF CONFERENCING SERVICE DEVELOPMENT APIs

| REST Resource | Operation | HTTP action and resource URI | Request body parameters | Most important info in response |
|---|---|---|---|---|
| List of Conferences | Create conference | POST: /conferences | Conference model, media, floor control, technology, conference size, QoS requirements, etc. | ID and URI of created conference |
| Participant | Add participant | POST: /conference/ {conferenceId}/ participants | Participant description: name, URI | ID and URI of new participant |
| Floor | Add floor | POST: /conferences /{conferenceId} /floors | Floor description: chair, floor participants | ID and URI of new floor created |
| Subconference | Remove subconference | DELETE: /conferences /{conferenceId} /subconferences /{subconferenceId} | None | Success or failure indication |

TABLE II. CATEGORIZATION OF API PARAMETERS

| | Categories | Example Values | | |
|---|---|---|---|---|
| **Mandatory Aspects** | Conference Model | Pre-arranged conference | Dial-in conference, Dial-out conference | |
| | | Ad-hoc conference | | |
| | *Media* | At least one of audio, video and text | | |
| | Conferencing Technology | SIP-based | Signaling protocol | SIP by default. No need to specify. |
| | | | *Audio and video encodings* | No mandatory encodings. So, must specify. |
| | | WebRTC-based | Signaling protocol | No mandatory protocol. So, must specify. |
| | | | *Audio encodings* | Mandatory: G.711 and Opus. Can specify additional. |
| | | | *Video encodings* | Mandatory: H.264 and VP8. Can specify additional. |
| | | Hybrid (SIP-based + WebRTC-based) | Mandatory protocols and encodings from both technologies apply. Can specify additional. | |
| **Optional Aspects** | Floor control | At least one floor control policy, e.g., chair-moderated and round-robin. | | |
| | *Subconference* | Enabled or not | | |

It is assumed that the service enables private text chat after 30 minutes. Using a regular timer function (available in most programming languages), the service invokes another API to add instant messaging to the conference for 5 minutes. *Conference Orchestration and Management* selects IaaS C, activates a substrate and modifies conference to add instant messaging. On the new substrate, individual conference is created for 5 minutes and existing participants are added. Then, participants can start exchanging text messages.

IV. IMPLEMENTATION AND MEASUREMENTS

A. Software Architecture

Fig. 4 shows the software architecture. To evaluate the proof-of-concept of conferencing PaaS, a stripped down conferencing IaaS is also designed. Its software architecture is shown in the very same Fig. 4.

*1) Conferencing PaaS*

In Conferencing PaaS GUIs and APIs, *Service Development APIs* can be provided as a programming library (e.g., JAR file in Java and NPM module in JavaScript). For service deployment and management, *Service Management GUI* is used. *Substrates Support Management GUI* is used by the conferencing platform provider to manage substrate information. Management (Services and PaaS) component and Service Hosting and Execution component are reused from conventional PaaS architectures. So, we do not discuss these in detail here.

In Conference Orchestration and Management component, *Substrate Selector* chooses the most suitable conferencing IaaS, given the substrate requirements. *Substrate Orchestration Engine* composes the selected substrates into a full-fledged conference. *Conference Execution Engine* hosts the conferences. *Conference Scaling Decision Maker* monitors running conferences and requests scaling when needed. *Conference Manager* receives requests from northbound component and coordinates other sub-components to serve the requests. Conferencing IaaS Handler component communicates with the conferencing IaaSs using PaaS/IaaS interfaces.

*2) Conferencing IaaS*

Conferencing IaaS has two main components. The first one, *IaaS Manager,* communicates with PaaS and handles all incoming requests. Moreover, it has control on resources (e.g., RAM, HDD and CPU) allocated to substrates. It also does regular IaaS tasks such as SLA management and IaaS governance. The second component, *Substrate Manager*, instantiates the requested substrate and configures it based on the PaaS requirements.

B. Prototype

The prototype scenario includes a service provider offering dial-in audio conferencing service and a game application consuming that service. It also includes the conferencing PaaS and two conferencing IaaSs – both providing dial-in signaling and audio mixer substrates. Two use cases are considered: One selects substrates from the same IaaS and the other chooses substrates from different IaaSs.

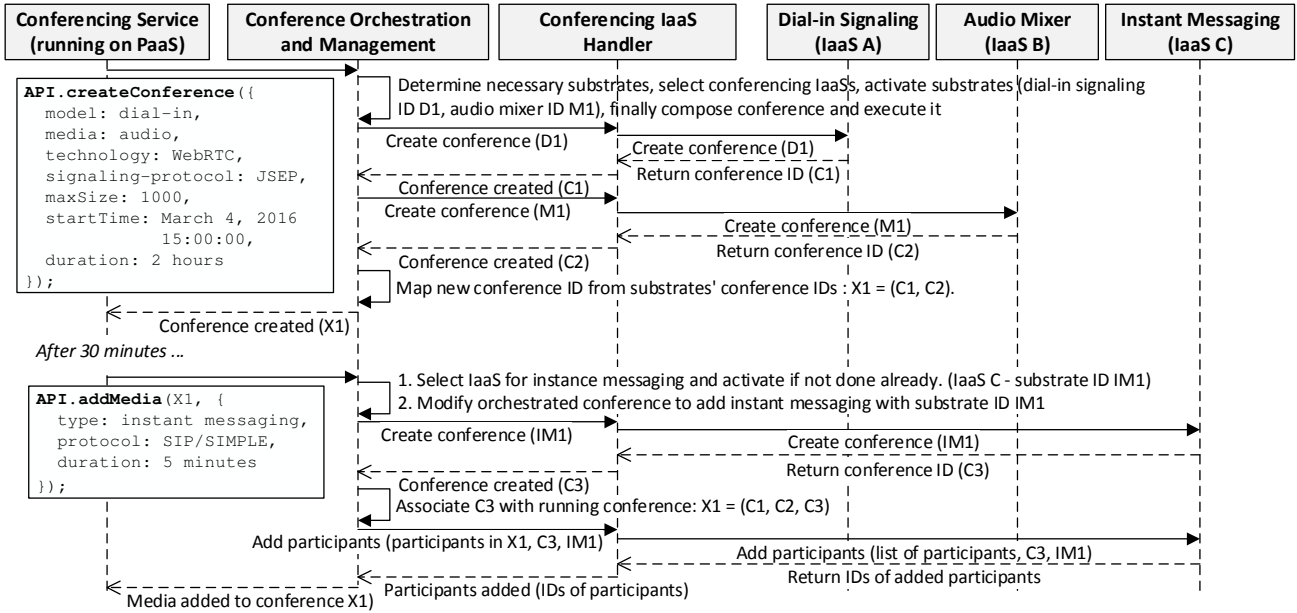

Fig. 3. Conference Creation

Cloud Foundry PaaS is extended, providing the implementation of typical PaaS components. For *Substrate Orchestration Engine* and *Conference Execution Engine*, open-source Camunda tool [16] is reused. *Conference Manager* and *Conferencing IaaS Handler* are implemented using Express.js framework. Advanced REST Client [15] is used to simulate SaaS API invocations by the game.

For conferencing IaaS, OpenStack [17] is reused. *Controller* is implemented as a custom Java application with REST-based APIs to communicate with the PaaS. For signaling and media handling substrates, Asterisk [18] is used as an open source framework. It is deployed on a machine with 4 GB RAM and two vCPUs running Ubuntu 14.04 LTS.

C. *Measurements*

Three scenarios are considered: (i) Non-cloud conferencing (NCC) where resources are allocated beforehand. The two other scenarios concern cloud-based conferencing, where conferencing PaaS is leveraged: (ii) Cloud single IaaS provider (CSIP) – PaaS selects required substrates for a conference from one IaaS, and (iii) cloud multiple IaaS provider (CMIP) – PaaS chooses substrates from different IaaSs. In CSIP, IaaS is assumed to host all substrates for a conference on the same VM.

The following three metrics are used: (i) Conference start time – i.e. the time required to get a conference ready upon the receipt of a request, (ii) participant joining time – i.e. the time required to add a participant to a running conference, and (iii) resource allocation – i.e. the total amount of allocated resources, such as RAM and CPU, to accommodate all participants. The last metric pertains to cloud-based scenarios.

As shown in Fig. 5(a), NCC takes the least time to start a new conference due to the absence of virtualization overhead. Since substrates need to connect over network in CMIP, it takes more time than in CSIP. Participant joining time is the least in NCC as shown in Fig. 5(b). Cloud-based scenarios take more time because of the notification overhead between IaaSs, PaaS and the game server. However, this is a one-time operation for a participant and does not contribute to the participant's communication delay. Moreover, based on

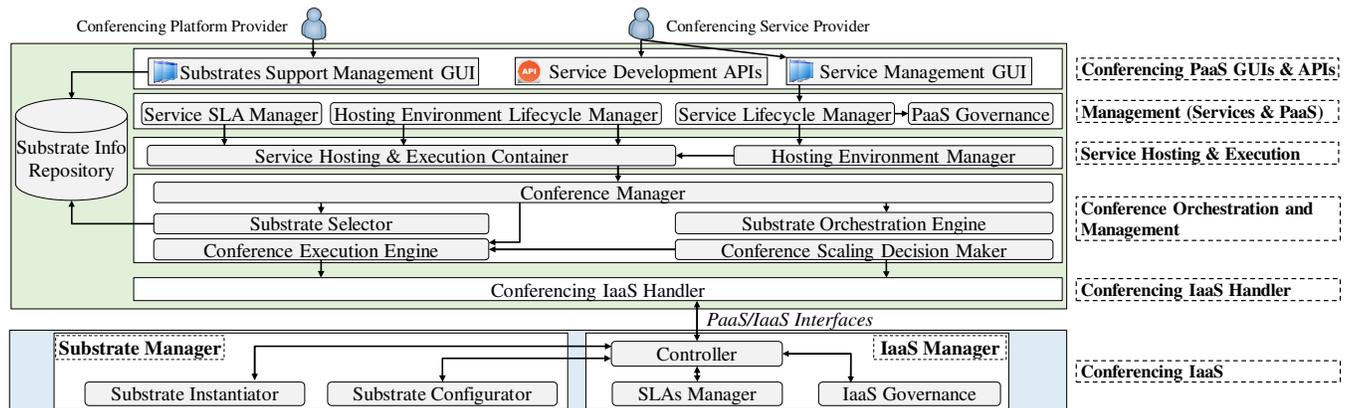

Fig. 4. Software Architecture

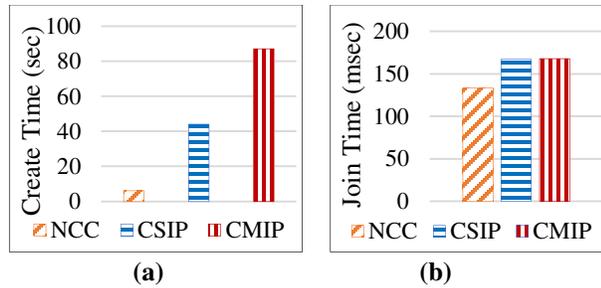

Fig. 5. (a) Average Conference Create Time (b) Participant Joining Time

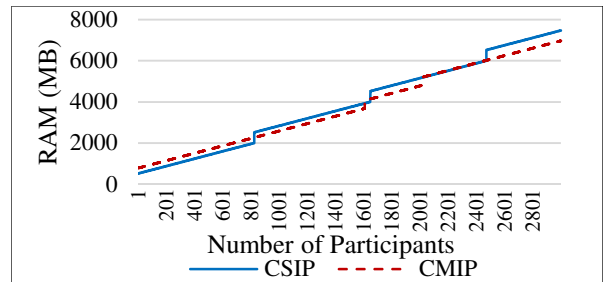

Fig. 6 Resource Allocation Evaluation

International Telecommunication Union (ITU) standards, this time is acceptable as long as it is below 400 msec [19]. Participant joining time of the two cloud-based scenarios are close as IaaSs can notify PaaS in parallel. Thus, the proposed architecture satisfies the QoS requirement.

Although in cloud-based scenarios, the start time and the participant joining time are more than those in NCC, it helps to achieve resource efficiency and reduce costs. Fig. 6 shows the allocated amount of RAM for a conference with between 1 and 3000 participants. To simulate conference scaling, conference size is increased by 200 participants every 10 minutes. The results are based on the observed resource usage per participant. IaaSs are assumed to scale up and out VMs while maintaining QoS requirements. In NCC, there are always some idle and non-utilized resources because of upfront resource provisioning. Hence, it is not shown in Fig. 6. CSIP scales better than CMIP (i.e. allocates less resources) for smaller conferences whereas CMIP wins for bigger conferences, because in CMIP, substrates are hosted on separate VMs as they are chosen from different IaaSs. For smaller conferences, it leads to more VMs and more non-utilizable resources (e.g., resources consumed by operating system) than in CSIP. However, with the increase of conference size, CMIP achieves better scalability because of the less VMs and more utilizable resources than in CSIP.

## V. CONCLUSION

A novel conferencing PaaS architecture for service providers is proposed, to easily provision conferencing services. The proposed PaaS cooperates with conferencing IaaSs to scale conferences elastically. The experiments show that cloud-based conferencing service provisioning can provide better resource efficiency. Several future algorithmic works have been identified, for instance, algorithms for PaaS to scale conference in terms of the number of participants and algorithms to select the most suitable conferencing IaaS, given the requirements of a substrate.

## VI. ACKNOWLEDGMENT

This work is supported in part by an NSERC/Ericsson CRD grant, the NSERC SAVI Research Network, and an NSERC Discovery grant.

## VII. REFERENCES

[1] L. M. Vaquero, L. Rodero-Merino, J. Caceres, and M. Lindner, "A break in the clouds: towards a cloud definition," *ACM SIGCOMM Comput. Commun. Rev.*, vol. 39, no. 1, pp. 50–55, 2008.
[2] R. H. Glitho, "Cloud-based multimedia conferencing: Business model, research agenda, state-of-the-art," in *Commerce and Enterprise Computing (CEC), 2011 IEEE 13th Conference on*, 2011, pp. 226–230.
[3] M. Jacobs and P. Leydekkers, "Specification of synchronization in multimedia conferencing services using the TINA lifecycle model," *Distrib. Syst. Eng.*, vol. 3, no. 3, p. 185, 1996.
[4] J. Li, R. Guo, and X. Zhang, "Study on service-oriented Cloud conferencing," in *Computer Science and Information Technology (ICCSIT), 2010 3rd IEEE International Conference on*, 2010, vol. 6, pp. 21–25.
[5] P. Rodríguez, D. Gallego, J. Cerviño, F. Escribano, J. Quemada, and J. Salvachúa, "Vaas: Videoconference as a service," in *Collaborative Computing: Networking, Applications and Worksharing, 2009. CollaborateCom 2009. 5th International Conference on*, 2009, pp. 1–11.
[6] F. Taheri, J. George, F. Belqasmi, N. Kara, and R. Glitho, "A cloud infrastructure for scalable and elastic multimedia conferencing applications," in *Network and Service Management (CNSM), 2014 10th International Conference on*, 2014, pp. 292–295.
[7] Y. Feng, B. Li, and B. Li, "Airlift: Video conferencing as a cloud service using inter-datacenter networks," in *Network Protocols (ICNP), 2012 20th IEEE International Conference on*, 2012, pp. 1–11.
[8] R. Cheng, W. Wu, Y. Lou, and Y. Chen, "A cloud-based transcoding framework for real-time mobile video conferencing system," in *Mobile Cloud Computing, Services, and Engineering (MobileCloud), 2014 2nd IEEE International Conference on*, 2014, pp. 236–245.
[9] J. Liao, C. Yuan, W. Zhu, P. Chou, and others, "Virtual mixer: Real-time audio mixing across clients and the cloud for multiparty conferencing," in *Acoustics, Speech and Signal Processing (ICASSP), 2012 IEEE International Conference on*, 2012, pp. 2321–2324.
[10] C. Vecchiola, X. Chu, and R. Buyya, "Aneka: a software platform for .NET-based cloud computing," *High Speed Large Scale Sci. Comput.*, vol. 18, pp. 267–295, 2009.
[11] "Cloud Foundry Overview." [Online]. Available: http://docs.cloudfoundry.org/concepts/overview.html. [Accessed: 04-Nov-2015].
[12] Q. Z. Sheng, X. Qiao, A. V. Vasilakos, C. Szabo, S. Bourne, and X. Xu, "Web services composition: A decade's overview," *Inf. Sci.*, vol. 280, pp. 218–238, 2014.
[13] T. Yu and K.-J. Lin, "Service Selection Algorithms for Composing Complex Services with Multiple QoS Constraints," in *Service-Oriented Computing - ICSOC 2005*, B. Benatallah, F. Casati, and P. Traverso, Eds. Springer Berlin Heidelberg, 2005, pp. 130–143.
[14] M. Garriga, C. Mateos, A. Flores, A. Cechich, and A. Zunino, "RESTful service composition at a glance: A survey," *J. Netw. Comput. Appl.*, vol. 60, pp. 32–53, 2016.
[15] "Advanced REST client." [Online]. Available: https://chrome.google.com/webstore/detail/advanced-rest-client/hgmloofddffdnphfgcellkdfbfbjeloo. [Accessed: 27-Feb-2016].
[16] "camunda BPM." [Online]. Available: https://github.com/camunda. [Accessed: 27-Feb-2016].
[17] "Home » OpenStack Open Source Cloud Computing Software." [Online]. Available: https://www.openstack.org/. [Accessed: 26-Feb-2016].
[18] "Asterisk.org." [Online]. Available: http://www.asterisk.org/. [Accessed: 11-Sep-2015].
[19] O. T. Time, "ITU-T Recommendation G. 114," *ITU-T May*, 2000.